\documentstyle[twocolumn,prl,aps]{revtex}

\begin{document}

\draft

\title{Theory of Ferromagnetic Superconductivity}

\author{Kazushige Machida and Tetsuo Ohmi$^{1}$}
\address{Theoretische Physik, ETH-H\"{o}nggerberg,
CH-8093 Z\"{u}rich, Switzerland}
\address{$^{\dagger}$Department of Physics, Okayama University,
         Okayama 700-8530, Japan}
\address{$^{1}$Department of Physics, Graduate School of Science, 
Kyoto University, Kyoto 606-8502, Japan}
\date{\today}

\maketitle

\begin{abstract}
It is argued that the pairing symmetry realized in a ferromagnetic superconductor UGe$_2$ must be a non-unitary triplet  pairing.
This particular state is free from the Pauli limitation and can survive under a huge internal molecular filed. To check our identification we examine its basic properties and several experiments are proposed. In particular, the external field is used to raise $T_c$ by controlling the internal spontaneous dipole field.
\end{abstract}


\narrowtext


Ferromagnetism (FM) and superconductivity (SC) are thought to
be basically mutually repulsive.
Since Ginzburg\cite{ginzburg} points out a possibility of its coexistence 
under the condition that the magnetization is less than
the thermodynamic critical field,
there are many experimental investigations performed, 
starting by Matthias et al\cite{matthias} who consider impurity ferromagnetism
in a superconductor.
Although the coexistence between antiferromagnetism
and superconductivity are rather easy to realize and actually observed 
in several compounds\cite{maple}
because the antiferromagnetic moments spatially averaged 
over the SC coherent length vanish, the
ferromagnetic case is difficult to realize.
Rare exceptions of this case are rare-earth ternary compounds
HoMo$_6$S$_8$ and ErRh$_4$B$_4$ where in a narrow temperature region just
below the Curie temperature $T_{FM}$
the coexistence of FM and SC is attained\cite{note}. 
When the rare-earth 4f moments completely
align at lower $T$, SC is wiped out by a strong internal field.
So far there is no known SC which can fully sustain such a large molecular field.

As for the theoretical developments, Anderson and Suhl\cite{anderson}
discuss the cryptoferromagnetism as a possible coexistence phase
where the FM is modified to a long-period modulated spin structure. 
This modified RKKY interaction is mediated by superconducting pairs.
Similarly to this idea, Blount and Varma\cite{blount} also propose a magnetic 
spiral phase coexisting with SC to gain electromagnetic dipole interaction
between localized moments.

The recent discovered material UGe$_2$\cite{saxena} seems to be difficult to
understand in terms of these theories so far proposed and seems to require
a novel concept to interpret the data:
Because previous theories assume that two groups of
electrons are distinguishable and clearly separated spatially.
FM comes from the well localized 4f electrons while
SC pairs are formed by conduction electrons. Here the situation
is much more intricate; we cannot separate two groups in a well
defined way from the outset, where 5f electrons from U-atoms 
play double roles both for FM and SC. It is a quite 
interesting problem how to describe its double roles played 
by 5f electrons microscopically,
here we confine our discussions to a phenomenological level
without going into pairing mechanism, which is closely related to it.

At ambient pressure UGe$_2$ is an itinerant metallic ferromagnet whose 
Curie temperature $T_{FM}$=52K and the spontaneous moment $\sim 1.4\mu_B$
/U-atom.
The easy axis is the $a$ axis in orthorhombic crystal\cite{onuki}.
Upon increasing pressure $P$, both $T_{FM}$ and the spontaneous moment
moderately decrease gradually and at $P\sim$1GPa the SC starts to 
appear and the SC transition temperature $T_c$ increases,
taking a maximum $T_c\sim 0.8$K above which FM already 
sets in at a high $T$ of $T_{FM}\sim 30$K
whose moment $\sim 1\mu_B$/U-atom.
Further increasing $P$, both $T_{FM}$ and $T_c$ drop down to disappear
almost simultaneously around $P\sim$1.7GPa.
Thus the SC region is entirely covered by FM. The SC is confirmed to
coexist with FM by neutron experiment\cite{huxley}.
This phenomenon is difficult to understand in 
terms of the previous framework based in a conventional singlet pairing state.
The internal ferromagnetic molecular field coming from the ordered
magnetic system through the exchange interaction amounts to an
order of few hundred tesla (T) in view of the exchange splitting
$\sim$70meV for the up spin and down spin Fermi surfaces\cite{yamagami}.
According to Tsutsui et al\cite{tsutsui} the hyperfine field probed by
U M{\"o}ssbauer specroscopy is $\sim$240T.
This huge exchange field apparently excludes not only any singlet
pairing category, but also certain forms of triplet pairing category,
namely, unitary triplet states. These are all limited by the Pauli 
paramagnetic field $H_p\sim$1.3$T_c$(T). Anderson and Suhl\cite{anderson},
and Blount and Varma\cite{blount}
theories are not applicable here, and Fulde and Ferrell (FFLO)
state only slightly
enhances $H_{c2}$ by at most 10\%\cite{saint},
but definitely not possible for $>$100T.

In fact, only possible pairing symmetry under such a strong internal
field is non-unitary triplet state, which is free from the Pauli limit
\cite{ohmi}.
Thus, we investigate this possibility in light of the present
material UGe$_2$, examine the available data and predict some of the 
interesting phenomena associated with this non-unitarity\cite{sugiyama}.

A non-unitary triplet state is described by the order
parameter ${\hat\Delta}({\bf k})=i({\bf d}({\bf k})
\cdot{\vec \sigma})\sigma_y$
in a 2$\times$2 matrix form. The three dimensional complex vector 
${\bf d}(\bf k)$ fully characterizes the triplet pairing state. If  
${\bf d}({\bf k})$ is a complex number, the product 
${\hat\Delta}({\bf k}){\hat\Delta}({\bf k})^{\dagger}=
|{\bf d}({\bf k})|^2\sigma_0+i({\bf d}({\bf k})
\times{\bf d}^{*}({\bf k}))\cdot{\vec \sigma }$ is
not a multiple of unit matrix and ${\hat\Delta}({\bf k})$
becomes non-unitary. Thus in a non-unitary state time
reversal symmetry necessarily is broken spontaneously 
and spontaneous moment ${\bf m}({\bf k})\propto i{\bf d}({\bf k})
\times{\bf d}^{*}({\bf k})$ appears at each $\bf k$ point, 
yielding the macroscopic averaged moment $<{\bf m}({\bf k})>$
provided that the its Fermi surface average non-vanishes.
There are two kinds of non-unitary state among several possible
forms, depending on its
orbital structure; one is bipolar type and non-bipolar one.
In the former (latter) the real and imaginary parts 
of the complex ${\bf d}(\bf k)$
vector are ascribed by different (same) orbital function(s). The
average spontaneous moment in the latter (former) is nonvanishing (vanishing)
in general by symmetry.
Since the additional Zeeman magnetic energy is gained in the non-bipolar,
in the following we only consider the non-bipolar state.

We derive an appropriate Ginzburg-Landau (GL)
free energy density functional allowed in the presence 
of the FM described in
terms of the complex order parameter ${\vec \eta}$
of the coefficients of the $d$ vector.
The macroscopic ferromagnetic
order parameter ${\bf M}$ is proportional to the magnetization
which is estimated as $\sim$0.2T in the ambient pressure\cite{saxena}.
The complex order parameter ${\vec \eta}$ is coupled linearly and quadratically 
with ${\bf M}$, namely, $i\gamma{\bf M}
\cdot{\vec \eta}\times{\vec \eta}^{*}$ and $M^2
{\vec \eta}\cdot{\vec \eta}^{*}$ respectively.
Thus it is convenient to write ${\vec \eta}=(\eta_x,\eta_y,0)$
or $\eta_{\pm}={1\over \sqrt2}(\eta_x\pm i\eta_y)$ by
choosing ${\bf M}=(0,0,M)$, implying that the Cooper pair spin orientation
points to the ${\bf M}$ direction. 
We choose a coordinate system: $x\parallel b,y\parallel c,z\parallel a$ where the magnetic 
easy axis is the $a$ axis. In view of the lattice constant $a$ and 
$c$ are almost same, only 2\% difference, we regard the crystal structure
as tetragonal where the $(a,c)$ forms the basal plane.

So far we have discussed the possible coupling terms with FM
allowed by symmetry in order to explicitly see the effects of the FM
formation. If we further consider the 
microscopic origin of SC, which appears under the Zeeman split Fermi 
surfaces (FS),
leading to the distinct transition temperatures and the different
fourth order terms for $\eta_{\pm}$ from the outset.
Thus we finally arrive at the following generic GL form appropriate
to UGe$_2$:

\begin{eqnarray} &&
f_{bulk}=\alpha_+|\eta_+|^2+\alpha_-|\eta_-|^2
\nonumber \\ &&
+{1\over 2}\beta_+|\eta_+|^4
+\beta|\eta_+|^2|\eta_-|^2+{1\over 2}\beta_-|\eta_-|^4
\label{eq:glbulk}
\end{eqnarray}

\noindent
where we introduce $\alpha_{\pm}=\alpha_0^{\pm}(T-T_{c}^{\pm})$
and $\beta_{\pm}>0$ for
the stability condition of the system. In the following 
$T_c^+$ is identified to the observed $T_c$
below which $\eta_+$ becomes non-vanishing.

It is easy to derive the condition for the second SC transition 
 to occur: $\alpha_0^-T_c^-/\alpha_0^+T_c^+>\beta/\beta_+$
at $T_{c2}=(\beta_+ \alpha_0^-T_c^--\beta\alpha_0^+ T_c^+)/(\beta_+ \alpha_0^--\beta\alpha_0^+)$
at which the remaining order parameter $\eta_-$ starts to appear.
Thus there is a chance to find the double transition near the 
critical pressure  1.7GPa where FM is suppressed and $T_c^-$
will approach $T_c^+$. We note that this transition is analogous to the 
A$_1$-A phase transition in superfluid $^3$He described by Ambegaokar
and Mermin\cite{ambegaokar}.
Near the critical pressure where $M$ is small, we can estimate relative changes of the two transitions by expanding the density of 
states $N(\epsilon_F)$ at $\epsilon_F$, giving rise to 
$T_{c1,2}=T_{c0}(1\pm M{N'(\epsilon_F)\over N(\epsilon_F)}\ln{\omega\over T_{c0}})$.
The derivative of the densities of states $N'(\epsilon_F)$
and the energy cutoff $\omega$.

The quasi-particle excitation spectrum for a non-bipolar type 
${\bf d}({\bf k})={\vec \eta}\phi({\bf k})$,
where the spin ${\vec \eta}$ and the orbital $\phi({\bf k})$
parts are separable, has two branches: 
$E_{\sigma}({\bf k})=\sqrt{\epsilon_{\sigma}^2({\bf k})
+\Delta_{\sigma}^2({\bf k})}$
$(\sigma=\pm)$. The gap functions are given
by $\Delta_{\pm}({\bf k})
=|\eta_x\pm\eta_y|\phi({\bf k})=|\eta_{\pm}|\phi({\bf k})$.
Thus one of the branches, say, $\Delta_-({\bf k})$
vanishes identically. That means that on the spin down Fermi
surface there is no superconducting gap formed, leaving 
it normal.

The nodal structure associated with non-unitarity leads to 
several observable predictions without specifying particular 
orbital function $\phi(\bf k)$. The total density of states
is given by $N(E)=N_+(E)+N_-(E)$ where 
$N_{\sigma}(E)=\Sigma_{\bf k}\delta(E-E_{\sigma}(\bf k))$.
At the SC transition temperature, only the spin up FS opens
the gap, thus the jump of the specific heat is substantially reduced 
from the BCS value (1.43). In the lowest $T$ the specific heat becomes
$C(T)=\gamma_-T$ where $\gamma_-$ is the densities of states 
at the FS for the spin down band.
According to an estimate by the spin polarized band structure calculation
\cite{yamagami} $\gamma_-$ is
substantial compared with the total $\gamma$ (=$\gamma_+$+$\gamma_-$).
Thus, it is quite observable. The existence of the $\gamma_-$
term governs the lowest $T$ thermodynamics, such as a $T$-linear term
in thermalconductivity and 
the $I-V$ characteristics of quasi-particle tunneling between 
a normal metal and UGe$_2$,  and etc.

We notice that if the second transition $T_{c2}$ really takes place
at lower $T$, there is no residual $\gamma_-$ contribution to 
the thermodynamic quantities mentioned above since the remaining spin down FS is also gapped by $\eta_{-}$.

As for the spin susceptibility $\chi_i$ ($i=a,b$ and $c$) 
probed by the Knight shift experiment, we 
naively expect that $\chi_b$ and $\chi_c$ ($\chi_a$)
must decrease (unchanges) below $T_c$ because the ${\bf d}(\bf k)$
 vector lies in the 
$(b,c)$ plane, but remain a finite value, corresponding to $\gamma_-$.
Note that $\chi_b$ and $\chi_c$ may not decrease appreciably
below $T_c$ because these transverse susceptibilities in FM
are known to be determined by the whole bands and insensitive to the
gap formation near FS.
The direction of the ${\bf d}(\bf k)$ vector is strongly locked to ${\bf M}//a$
where UGe$_2$ is known to have strong easy axis type of
the magnetic anisotropy and the magnetization is Ising-like\cite{saxena}.

Having analyzed the basic thermodynamic properties of non-unitary state
for UGe$_2$, we now proceed to examining electromagnetic
properties when the external field 
is applied, regarding UGe$_2$ a conventional type II superconductor.
We note, however, that without $H$ vortices with each having unit quantum flux are 
spontaneously created because the internal field $\sim$0.2T far exceeds 
the usual $H_{c1}$. In this sense the system lacks the complete Meissner phase 
in $H$ vs $T$ plane.
Let us introduce the free energy density $f_{grad}$
related to the magnetic field $H$

\begin{eqnarray} &&
f_{grad}=K_1\Sigma_{\sigma=\pm}\Sigma_{j=y,z}|D_j\eta_{\sigma}|^2
+K_2\Sigma_{\sigma=\pm}|D_x\eta_{\sigma}|^2
\label{eq:glgrad}
\end{eqnarray}

\noindent
where $D_j=\partial_j-i{2e\over \hbar c}A_j$ and the unit flux
$\phi_0={\hbar c\over 2e}$. We have assumed a simple 
orbital function $\phi({\bf k})$
which does not break tetragonal symmetry in the
basal plane $(a, c)$ or $(y,z)$ for simplicity.
We reserve its extension to a future work when the orbital function
turns out to be more complex. The essence of the following arguments is not
altered. Near $T_c$ when $H$ is applied parallel to the $z$ axis (the $a$ axis)
the above is written as 

\begin{eqnarray}
f_{grad}=K_1({d\eta_+\over dx})^2+({2\pi\over \phi_0})^2K_2(M+\mu_0H)^2x^2\eta_+
\end{eqnarray}

\noindent
because the magnetic induction ${\bf B}=(0,0,\mu_0H+M)$.
The minimization of the free energy

\begin{eqnarray}
{2\pi\over \phi_0}\sqrt {K_1K_2}|M+\mu_0H^a_{c2}|=\alpha_0(T_{c}-T)
\end{eqnarray}

\noindent
readily yields the upper
critical field $H_{c2}^a$ as

\begin{eqnarray}
\mu_0H_{c2}^a={\phi_0\over 2\pi}{\alpha_0\over \sqrt {K_1K_2}}(T_c-T)
\end{eqnarray}

\noindent
where the transition temperature is redefined as
$T_c-{2\pi\over \phi_0}{\sqrt{K_1K_2}\over\alpha_0}M\rightarrow T_c$ and $\alpha_0^+\rightarrow\alpha_0$.
When $H\parallel y(\parallel c)$, by a similar way we obtain

\begin{eqnarray}
f_{grad}=K_2({d\eta_+\over dx})^2+
({2\pi\over \phi_0})^2K_1\{M^2+(\mu_0H)^2\}x^2\eta_+
\end{eqnarray}

\noindent
which yields 

\begin{eqnarray}
{2\pi\over \phi_0}\sqrt {K_1K_2}\sqrt{M^2+(\mu_0H^b_{c2})^2}
=\alpha_0(T_{c}-T)
\end{eqnarray}

\noindent
or, approximately near $T_c$
\begin{eqnarray}
\mu_0H_{c2}^c\sim{\sqrt{2M}}{\sqrt {\mu_0H_{c2}^a}}.
\end{eqnarray}

\noindent
This means that near $T_{c}$, $H_{c2}^c$ exhibits a root singularity
as a function of $T$ with the slope being infinite. This is also
true for $H//x(//b)$, namely,

\begin{eqnarray}
H_{c2}^b\sim{\sqrt{K_2\over K_1}}H_{c2}^c
\end{eqnarray}

\noindent
where it shows not only a root singular behavior at $T_c$,
but also is scaled with $H_{c2}^c$ by a constant factor.
The large slope in $H_{c2}^b$ is observed in a certain
pressure region of UGe$_2$ by Huxley\cite{private}

In the basal $(a,c)$ plane the angular dependence of $H_{c2}(\theta)$
($\theta$ is the angle from the $a$ axis) is calculated as 

\begin{eqnarray}
\{1-({H_{c2}^a\over H_{c2}^c})^2\}\cos\theta{H_{c2}(\theta)\over H_{c2}^a}
+({H_{c2}(\theta)\over H_{c2}^c})^2=1.
\end{eqnarray}

\noindent
by a simple arithmetics. Since the system is polar axis symmetric, namely
the magnetization is of vectorial nature, $M>0$ differs from $M<0$.
Thus when external field is reversed from $M>0$ to $M<0$ direction,
$H_{c2}^{-a}$ behaves differently. Namely 
provided that the magnetization stays its original orientation 
under the reversed field which is $H\lesssim 300$G, judging from the magnetization
curve (Fig. 1 in Ref. \cite{saxena}) $H_{c2}^{-a}$ 
is  given from eq. (4)

\begin{eqnarray}
\mu_0H_{c2}^{-a}={\phi_0\over 2\pi}{\alpha_0\over \sqrt {K_1K_2}}(T-T_c)
\end{eqnarray}

\noindent
for $T>T_c$.
In this situation $T_c$ goes up by 30mK$\sim$40mK. 
Then on further increasing field as the magnetization is quickly
reversed from its original direction and begins decreasing, the $H_{c2}^{-a}$
curve changes into original $H_{c2}^{a}$ curve. Thus it is expected that the 
SC state is reentrant by increasing $H$ under a fixed $T$ just above $T_c$.
This $T_c$ rise and the associated reentrant phenomenon is deeply rooted 
in the fact that the SC survives under the influence of the ferromagnetic state. Upon controlling FM by external field we can
manipulate and raise $T_c$. Physically the $T_c$ rise occurs because the external
field can cancel the spontaneously building internal field
due to the ferromagnetic polarization which is to lower the hypothetical transition temperature and the magnetic induction $\bf B$ can become smaller
than the internal field $M$.

The possible orbital forms in triplet pairing case allowed in 
tetragonal symmetry are given by Volovik and Gorkov\cite{volovik}
for strong spin-orbit coupling case. Since SC appears in the presence
of FM, the spin part of the order parameter is limited to the 
form: $\hat{\bf b}+i\hat{\bf c}$ ($\hat{\bf a}$, $\hat{\bf b}$
and $\hat{\bf c}$ are unit vectors along the three crystal axes).
In the strong spin orbit coupling case
all the one-dimensional representations (A$_{1u}$, A$_{2u}$, B$_{1u}$
and B$_{2u}$) are excluded, only the pairing function 
${\hat k_a}(\hat{\bf b}+i\hat{\bf c})$
remains as a candidate among the two-dimensional representation E$_{u}$
in their classification.
Since the spin orbit coupling is not the largest energy scale in
the present ferromagnetic superconductor, the weak spin orbit 
coupling scheme shown next is more relevant classification scheme.

In the weak spin-orbit case\cite{ozaki}, which was the case for UPt$_3$ where
another exotic superconducting pairing state (either non-unitary 
bipolar state similar to the present material or the triplet
planar state) is realized\cite{machida} there are several
possible orbital functions allowed coupled to the spin function 
$\hat{\bf b}+i\hat{\bf c}$. Namely, $k_ak_ck_b(k_a^2-k_c^2)$ (A$_{1u}$),
$k_b$ (A$_{2u}$), $k_ak_ck_b$ (B$_{1u}$) and $k_b(k_a^2-k_c^2)$ (B$_{2u}$)
as the one-dimensional representations with $k_i (i=a,b,c)$ 
being the unit vector for
reciprocal space. As for the two-dimensional representation E$_u$,
the orbital functions of the forms are allowed:
$\lambda_1(k)$, $\lambda_1(k)+\lambda_2(k)$ and $\lambda_1(k)+i\lambda_2(k)$
are listed where $\lambda_1(k)=k_a$ and $\lambda_2(k)=k_c$, or 
$\lambda_1(k)=k_ak_b^2$ and $\lambda_2(k)=k_ck_b^2$.
The bipolar type pairing state $\lambda_1(k)\hat {\bf a}
+i\lambda_2(k)\hat {\bf c}$
is excluded. If we take literally the orthorhombic crystal symmetry,
the strong spin-orbit classification does not give rise to a suitable
spin function of $\hat{\bf b}\pm i\hat{\bf c}$ among the four one-dimensional representations such as 
$k_a\hat {\bf a}$, $k_c\hat {\bf c}$ (A$_{1u}$), 
and $k_a\hat {\bf c}$, $k_c\hat {\bf a}$ (B$_{1u}$)
and its linear combinations. On the other hand, 
in the weak spin-orbit coupling 
case where there is no two-dimensional representation all one-dimensional 
presentations have  suitable orbital functions attached to 
the spin part $\hat {\bf b}+i\hat {\bf c}$, namely, $k_ak_ck_b$ (A$_{1u}$), 
$k_b$ (B$_{1u}$), $k_c$ (B$_{2u}$) and $k_a$ (B$_{3u}$).
It is interesting to note that $(\lambda_1(k)+i\lambda_2(k))
(\hat{\bf b}+i\hat {\bf c})$
has not only the spin angular moment but also the orbital angular moment,
both pointing to the ferromagnetic moment direction.
The allowed states have line node(s) in general except for 
$(k_a+ik_c)(\hat{\bf b}+i\hat {\bf c})$  in the two-dimensional representation if the up-spin Fermi surface opens along the $b$ axis as predicted 
by band structure calculation\cite{yamagami}.

We can design several interesting experiments associated with 
the coexistence between FM and SC in the zero-field cooling 
where FM domains are formed, consisting of $\bf M$ and $\bf -M$.
Each domain has the own SC order parameters, either 
$\hat{\bf b}+i\hat {\bf c}$, 
or  $\hat{\bf b}-i\hat {\bf c}$
in order to fit its spin angular moment with the
ferromagnetic moment direction. When one of the two domains, say, $\bf M$ 
with $\hat{\bf b}+i\hat {\bf c}$ percolates throughout 
a whole system, the macroscopic superconducting state establishes. In contrast, under the field cooling process 
whose field strength strong enough to make the whole system the 
single domain, say, $\bf M$, the SC takes place immediately right at $T_c$.

This reentrant phenomenon mentioned before
provides direct evidence for the
coexistence.
The reentrance behavior is indeed observed at 1.35GPa
at much higher field region ($\sim$ a few T) may be different origin 
from our prediction. However, it should be kept in mind that the
superconducting properties in UGe$_2$ show strong sensitivity of the measured
current density\cite{huxley}. It might be related to the domain formation associated with FM.
Toward the critical pressure $P=1.7$GPa the FM is greatly suppressed.
There the two long-range orders FM and SC truly compete each other
while around the optimal pressure $P\sim$1.2GPa $T_{FM}$ 
is much higher than $T_c$. Thus the relationship between FM and SC is 
changing upon increasing $P$. As a consequence it could be
possible to change the pairing symmetry from the non-unitary to unitary state. This is one of possible explanations of the observed reentrance phenomenon at $P$=1.35GPa.

We also remark that along the FM domain boundaries or the Bloch walls
where the magnetization direction rotates so as to bridge the two
oppositely polarized states $\pm {\bf M}$, the spontaneously induced spin current flows 
because when transforming $\hat {\bf b}+i\hat {\bf c}$ state into $\hat {\bf b}
-i\hat {\bf c}$ 
the SC phase also continuously changes.

We are grateful for useful discussions and information
with G. Lonzarich, A. Huxley and S. Saxena. We also acknowledge
useful discussions with M. Ozaki and H. Yamagami.


$^{\dagger}$ permanent address


\end{document}